\documentclass[12pt]{article}

\newcommand{\be}{\begin{equation}}
\newcommand{\ee}{\end{equation}}
\newcommand{\bea}{\begin{eqnarray}}
\newcommand{\eea}{\end{eqnarray}}

\newcommand{\gs}{\ensuremath{g_s}} 

\newcommand{\cN}{{\mathcal{N}}}

\newcommand{\aD}{\ensuremath{\overline{\mbox{D3}}}}
\newcommand{\DD}{\ensuremath{\mbox{D-}\overline{\mbox{D}}}}
\newcommand{\bN}{\ensuremath{\bar{N}}}
\newcommand{\mm}{\ensuremath{m_{\mathrm{FT}}}}
\newcommand{\mg}{\ensuremath{m_{\mathrm{SG}}}}
\newcommand{\sm}{\ensuremath{s_{\mathrm{FT}}}}
\newcommand{\sg}{\ensuremath{s_{\mathrm{SG}}}}
\newcommand{\jm}{\ensuremath{j_{\mathrm{FT}}}}
\newcommand{\jg}{\ensuremath{j_{\mathrm{SG}}}}
\newcommand{\qm}{\ensuremath{Q_{\mathrm{FT}}}}
\newcommand{\qg}{\ensuremath{Q_{\mathrm{SG}}}}

\title{The Entropy of the Rotating Charged Black Threebrane from a
 Brane-Antibrane System}
\author{Alberto G\"uijosa\footnote{e-mail: alberto@nuclecu.unam.mx}
\\{\small Departamento de F\'{\i}sica de Altas Energ\'{\i}as,
Instituto de Ciencias Nucleares}\\ {\small Universidad Nacional
Aut\'onoma de M\'exico}\\
{\small A.P. 70-543, M\'exico D.F. 04510}
 \and H\'ector H. Hern\'andez Hern\'andez\footnote{e-mail:
hehh@xanum.uam.mx.} \\ Hugo A. Morales T\'ecotl\footnote{e-mail:
hugo@xanum.uam.mx.
Associate member of ICTP, Trieste, Italy.} \\
{\small Departamento de F\'{\i}sica, Universidad Aut\'onoma
Metropolitana Iztapalapa}\\ {\small A.P. 55-534, M\'exico D.F.
09340, M\'exico}}
\date{}

\begin{document}
\maketitle
\begin{abstract}
We show that a model based on a D3-brane--anti-D3-brane system at
finite temperature, proposed previously as a microscopic
description of the non-rotating black threebrane of type IIB
supergravity arbitrarily far from extremality, can also
successfully reproduce the entropy of the rotating threebrane with
arbitrary charge (including the neutral case, which corresponds to
the Kerr black hole in seven dimensions). Our results appear to
confirm in particular the need for a peculiar condition on the
energy of the two gases involved in the model, whose physical
interpretation remains to be elucidated.
\end{abstract}
\section{Introduction}

The problem of finding, through explicit state-counting, a
statistical-mechanical interpretation of the Bekenstein-Hawking
formula for the entropy of a black hole has long been regarded as
a crucial test for any theory that aims at describing gravity
microscopically. In recent years, the two main approaches to
quantum gravity, string theory and loop quantum gravity, have given strong
indications that they can successfully overcome this test (the
original references are \cite{stringbh,loopbh}; for reviews, see,
e.g., \cite{stringrev,looprev}). In both approaches, however, there are
practical as well as conceptual issues that remain to be grappled
with.

In particular, most of the string-theoretic analyses, going back
to the pioneering work \cite{stringbh}, have concentrated on black
holes (or branes) that are either extremal or near-extremal. The
far-from-extremal regime, which includes in particular the neutral
(Schwarzschild) black hole, has proven to be more challenging. In
the past, it has been modelled in terms of a highly excited
fundamental string \cite{stringbhcorr} (as a particular instance
of the correspondence principle \cite{hp}), and in the matrix
theory \cite{bfss} context \cite{matrixbh}. In these two
approaches the microscopic state-counting leads to rough agreement
with the black hole entropy, a result that is  clearly
non-trivial, but falls short of providing a fully quantitative
account of the black hole thermodynamics.

A third approach, related to matrix theory, makes use of a boost
and various dualities to map the far-from-extremal configuration
onto a near-extremal one \cite{boostbh}, and is capable of
producing quantitative agreement. In light of the analysis of the
boost given in \cite{seiberg}, however, the validity of the
mapping procedure is not altogether clear (see for instance
\cite{rindep}).

A fourth approach invokes the gauge/gravity correspondence
\cite{malda,imsy} to relate a specific non-asymptotically-flat
black hole to the quantum mechanics of D0-branes. Remarkably, in
this context it is possible to extract directly from a mean-field
approximation to the strongly-coupled gauge theory not only the
form of the non-extremal supergravity entropy, but also geometric
properties such as the location of the horizon \cite{kll}. {}Based
on the results of this line of work, an effective description for
a large class of black holes (including the Schwarzschild case)
has been proposed, in terms of a gas of non-interacting
quasi-particles \cite{ikll} (a similar picture has been considered
in \cite{halyo}).

A couple of years ago it was shown that the entropy of the black
threebrane of Type IIB supergravity and the twobrane and fivebrane
of eleven-dimensional supergravity arbitrarily far from
extremality (including in particular the Schwarzschild black hole
in seven, nine, and six spacetime dimensions) can be reproduced
employing microscopic models based on branes and antibranes
\cite{beto1,beto2}. (A different model for the Schwarzschild black
hole, based on \emph{Euclidean} brane-antibrane pairs, has been
considered in \cite{krause}.) Already in \cite{hms} it had been
noted that the entropy of the non-extremal D1-D5 system could be
rewritten in a form suggestive of a model involving branes and
antibranes, an observation subsequently generalized to other
systems in \cite{antientropy}. The understanding of
D-brane--anti-D-brane systems at that time was however
insufficient to attempt a direct formulation of such a model. In
the past few years this situation has changed completely; starting
with \cite{ddbar} we have gained quantitative control over the
physics of \DD\ systems. Indeed, the model proposed in
\cite{beto1} emerged naturally from a study, carried out in the
same work, of the properties of \DD\ systems at finite temperature
(other such studies may be found in \cite{tempddbar}). The
brane-antibrane model is able to successfully account for various
properties of the black brane; in particular, the negative
specific heat and pressure of the system find a natural
explanation in terms of brane-antibrane annihilation.

In this paper we will subject the brane-antibrane model of
\cite{beto1} to an additional non-trivial test, by exploring the
possibility of describing with it the \emph{rotating} black
threebrane. The addition of rotation is interesting from the
theoretical perspective because it results in a significant
modification of the functional form of the supergravity entropy.
It is of course also of interest from the phenomenological
viewpoint, because it allows us to study the Kerr black hole, thus
bringing us closer to the actual black holes that are believed to
exist in our universe. Section \ref{entropysec} contains our
calculations, starting in  \ref{neutralsubsec} with a summary of
the model of \cite{beto1} and an analysis of the neutral rotating
threebrane, and  ending in \ref{chargedsubsec} with the
generalization to the case with arbitrary charge. Our conclusions
are presented in Section \ref{conclsec}.

\section{Entropy Determination} \label{entropysec}

We take from \cite{gubser1} (see also \cite{klt}) the rotating
black threebrane solution of Type IIB supergravity, with all but
one of the rotation parameters set to zero. The metric takes the
form
\bea
ds^2&=&{1\over\sqrt{f}}(-hdt^2+d\vec{x}^2)+\sqrt{f}\left[{dr^2\over\tilde{h}}
-{2lr_0^4\cosh\alpha\over r^4\Delta f}\sin^2\theta dtd\phi\right.
\nonumber\\
{}&{}&\qquad\qquad\qquad\qquad\qquad\qquad\left.+r^2(\Delta
d\theta^2 +\tilde{\Delta}\sin^2\theta d\phi^2 + \cos^2\theta
d\Omega_3^2)\right], \nonumber
\eea
with
\bea
f&=&1+\frac{r_0^4\sinh^2\alpha}{r^4\Delta}~,\nonumber\\
\Delta&=&1+\frac{l^2\cos^2\theta}{r^2}~,\nonumber\\
\tilde{\Delta}&=&1+{l^2\over r^2}+\frac{r_0^4
l^2\sin^2\theta}{r^6\Delta f}~, \nonumber\\
h&=&1-{r_0^4\over r^4\Delta}~, \nonumber\\
\tilde{h}&=&{1\over\Delta}\left(1+{l^2\over r^2}-{r_0^4\over
r^4}\right)~.\nonumber
\eea
This geometry has an ADM mass density
\be \label{mg}
\mg\equiv {M_{\mathrm{SG}}\over
V}={\pi^3\over\kappa^2}r_0^4\left(\cosh 2\alpha + {3\over
2}\right),
\ee
and an angular momentum density (conjugate to $\phi$)
\be \label{jg}
\jg\equiv {J_{\mathrm{SG}}\over V} = {\pi^3\over\kappa^2}l
r_0^4\cosh\alpha~.
\ee
There is an event horizon at
$r_H=\sqrt{\sqrt{r_0^4+l^4/4}-l^2/2}$, whose area corresponds to
an entropy density
\be \label{sg}
\sg\equiv {A_{H}/4G_{N}\over
V}={2\pi^2\over\kappa^2}r_0^4\sqrt{\sqrt{r_0^4+l^4/4}-l^2/2}\cosh\alpha~.
\ee
In addition, the solution involves a RR five-form, associated with
a charge
\be \label{qg}
\qg={\pi^{5/2}\over\kappa}r_0^4\sinh 2\alpha~.
\ee
\subsection{Neutral case}
\label{neutralsubsec}

As can be seen from (\ref{qg}), the neutral case corresponds to
setting $\alpha=0$. The mass (\ref{mg}) and angular momentum
(\ref{jg}) then reduce to
\begin{equation}
\mg=\frac{5\pi^3}{2\kappa^2} r_0^4, \qquad
\jg={\pi^3\over\kappa^2}l r_0^4~.
\end{equation}
Using these expressions in (\ref{sg}) we can eliminate $r_0$ and
$l$ in favor of $\mg$ and $\jg$, to obtain the supergravity
entropy as a function of the mass and angular momentum of the
brane,
\begin{equation}\label{sgneut}
\sg=\sqrt{\sqrt{2^9 5^{-5}\pi
\kappa^2\mg^5+4\pi^4\jg^4}-2\pi^2\jg^2}~.
\end{equation}
For small $\jg$, this implies
\begin{equation}
\label{sgneutapprox}
\sg=2^{9/4}5^{-5/4}\pi^{1/4}\kappa^{1/2}\mg^{5/4}\left[1-2^{-9/2}
5^{5/2}\pi^{3/2}{\jg^2\over \kappa \mg^{5/2}}+2^{-10}5^{5}\pi^3
{\jg^4\over \kappa^2 \mg^{5}}+ \mathcal{O}(\jg^6) \right].
\end{equation}

It was shown in \cite{beto1} that the first term in
(\ref{sgneutapprox}), the entropy for the non-rotating neutral
black threebrane, can be reproduced with a field-theoretic model
based on a system of $N$ D3-branes, $\bN=N$ anti-D3-branes, and
two gases of $\cN=4$ SYM particles, arising respectively from the
massless modes of 3-3 and $\overline{3}$-$\overline{3}$ open
strings. The total mass density of the system is then
\be \label{mmneut}
\mm=2N\tau_{3}+e~,
\ee
with $\tau_{3}=\sqrt{\pi}/\kappa$ the D3-brane tension and $e$ the
total energy (density) available to the gases. The entropy of the
system is entirely due to the $\cN=4$ SYM gases. In the regime of
interest, $\gs N\gg 1$ (where the supergravity solution is
reliable), SYM is strongly-coupled, and so the entropy of the two
gases cannot be determined perturbatively. It is however known
\cite{gkp,threequarters} via the AdS/CFT correspondence
\cite{malda},
\be \label{sm}
\sm=2^{9/4}3^{-3/4}\pi^{1/2}(e/2)^{3/4}\sqrt{N}~,
\ee
where we have taken into account the fact that each gas should
carry half of the available energy $e$. Equation (\ref{mmneut}) is
then used in (\ref{sm}) to eliminate $e$ in favor of $N$, and the
value of $N$ (the number of \DD\ pairs) determined by maximizing
$\sm$ at fixed $\mm$. This leads to
\begin{equation}\label{ne}
N=\frac{\mm}{5\tau_3}\qquad \Longleftrightarrow \qquad e={3\over
5}\mm~.
\end{equation}
As a consistency check, this gas energy corresponds to a
temperature $T\gg 1/\sqrt{\gs N}$, which as explained in
\cite{beto1} is precisely the condition for the \DD\ pairs not to
annihilate (i.e., for the effective tachyon potential to have a
\emph{minimum} at the open string vacuum).

Using (\ref{ne}) and (\ref{mmneut}) in (\ref{sm}), it was found in
\cite{beto1} that the entropy of the field-theoretic model exactly
coincides with that of the supergravity solution (the first term
in (\ref{sgneutapprox})), except for a factor of $2^{3/4}$. Given
the form of (\ref{sm}), this is equivalent to saying that the
supergravity entropy behaves as if each of the gases somehow had
access to the \emph{total} energy $e$, rather than the half,
$e/2$, that we would naturally assign to it. Remarkably, it was
found in \cite{beto1} that this same model could equally well
reproduce the entropy of the charged threebrane (where
$N\neq\bN$), and similar models (based respectively on
M2-$\overline{\mbox{M2}}$ and M5-$\overline{\mbox{M5}}$ pairs))
could account for the entropy of the twobrane and fivebrane
solutions of eleven-dimensional supergravity. In all cases there
was perfect agreement with the functional form of the entropy, and
a numerical discrepancy that could be resolved under the uniform
(albeit bizarre) assumption that each of the two gases carries in
some sense the total energy $e$.\footnote{In \cite{beto1} it was
noted that, to reproduce the functional form of the charged
threebrane entropy, one had to assume in any case that both gases
have the \emph{same} energy. There then remained the numerical
$2^{3/4}$ discrepancy. Clearly we do not need the equality of
energies as an additional assumption if we interpret the
supergravity expression for the entropy as implying that each gas
has access to the total energy.}

It is interesting then to test the microscopic model of
\cite{beto1} (and in particular, the peculiar associated condition
on the energies) by turning on a small amount $\jg$ of threebrane
angular momentum. In the field-theoretic description this
corresponds to considering SYM gases that carry a total R-symmetry
charge density $\jm=\jg$. The entropy of a charged $\cN=4$ SYM gas
at strong coupling is again predicted by the AdS/CFT
correspondence: for a given energy, charge, and gauge group size,
it is \cite{gubser1,klt}
\be\label{sgubser}
s(e,j,N)=\frac{2^{5/4}3^{-3/4}\sqrt{\pi
N}e^{3/4}}{\sqrt{\sqrt{1+\chi}+\sqrt{\chi}}}~, \qquad
\chi=\frac{27 \pi^2 j^4}{8N^2e^3}~.
\ee

In our microscopic model there are two gases, one on the D3-branes
and the other on the anti-D3-branes. Given that $N=\bN$, by
symmetry we expect each gas to contribute half the total charge,
$j=\jm/2$. For the energy one would expect a similar split; but,
in accordance with the findings of \cite{beto1} summarized above,
we instead take the energy of each gas to be the \emph{total}
energy $e$ given by (\ref{mmneut}). The entropy of our system is
thus
\begin{equation}\label{smn}
\sm=\frac{2^{9/4}3^{-3/4}\sqrt{\pi
N}e^{3/4}}{\sqrt{\sqrt{1+\chi}+\sqrt{\chi}}}~, \qquad
\chi=\frac{27 \pi^2 \jm^4}{2^7 N^2e^3}~.
\ee
For small $\jm$ this can be expanded into
\be \label{smnapprox}
\sm=2^{9/4}3^{-3/4}\sqrt{\pi N}e^{3/4} \left( 1-{\sqrt{\chi}\over
2}+ {\chi\over 8}+\ldots \right)~,
\ee
an expansion of the same form as the supergravity expression
(\ref{sgneutapprox}), which seems encouraging. To perform a
detailed comparison, we first need to fix the number of \DD\ pairs
as in \cite{beto1}, by maximizing the entropy with respect to $N$.
At this order, the equilibrium value of $N$ is found to be the
same as in the non-rotating case, Eq. (\ref{ne}). Using this and
(\ref{mmneut}) in (\ref{smnapprox}) we obtain
\begin{equation} \label{smneutapprox}
\sm=2^{9/4}5^{-5/4}\pi^{1/4}\kappa^{1/2}\mm^{5/4}\left[1-2^{-9/2}
5^{5/2}\pi^{3/2}{\jm^2\over \kappa \mm^{5/2}}+2^{-10}5^{5}\pi^3
{\jm^4\over \kappa^2 \mm^{5}}+ \mathcal{O}(\jm^6) \right]~,
\end{equation}
which correctly reproduces  the supergravity result
(\ref{sgneutapprox})~!

Having found agreement between supergravity and the microscopic
model up to order $j^{4}$, one is encouraged to determine whether
the agreement extends to higher order, perhaps even to the entire
expression (\ref{sgneut}). So we maximize (\ref{smn}) with respect
to $N$; remarkably, the equilibrium value of $N$ is again
(\ref{ne}). Plugging this back into (\ref{smn}), we obtain a
microscopic entropy
\be\label{smneut}
\sm=\frac{2^4 5^{-5/4}\pi^{1/4}\kappa \mm^{5/2}}{\sqrt{\sqrt{2^7
\kappa^2 \mm^5+5^5 \pi^3 \jm^4}+5^{5/2} \pi^{3/2}\jm^2}}~.
\ee
At first sight this appears to differ from the supergravity result
(\ref{sgneut}), but upon multiplying the numerator and denominator
by ${\sqrt{\sqrt{2^7 \kappa^2 \mm^5+5^5 \pi^3 \jm^4}-5^{5/2}
\pi^{3/2}\jm^2}}$, we obtain complete agreement between the two
expressions!

\subsection{Charged case}
\label{chargedsubsec}

Now we extend our analysis to the charged case, $\alpha\neq 0$. We
could in principle use again (\ref{mg}), (\ref{jg}) and (\ref{qg})
in (\ref{sg}) to obtain an expression for the supergravity entropy
as a function of the mass, angular momentum and charge of the
threebrane. It is however easier to keep $\alpha$, $r_0$ and $l$
in the analysis and attempt to reproduce (\ref{sg}) directly.

In the microscopic model we now have
\be \label{qm}
\qm=N-\bN\neq 0~,
\ee
and as before, the actual values of $N$ and $\bN$ should be
determined by maximizing the entropy. Given our experience with
the neutral case, it is natural to expect that the equilibrium
values of $N,\bN$ will again coincide with those found in
\cite{beto1} for the non-rotating system. We thus postulate (and
later check) that, at equilibrium,
\be \label{nnbar}
N={\pi^{5/2}\over 2\kappa}r_0^4 e^{2\alpha}, \qquad
\bN={\pi^{5/2}\over 2\kappa}r_0^4 e^{-2\alpha},
\ee
which is indeed the most natural way to split the supergravity
expression (\ref{qg}) into two parts. {}Comparing the mass of the
microscopic system,
\be \label{mmcharged}
\mm=(N+\bN)\tau_3+e~,
\ee
against the supergravity expression (\ref{mg}), we then infer that
(as in the non-rotating case) the energy of the gases should be
identified with
\be \label{echarged}
e={3\pi^3\over 2\kappa^2}r_0^4~.
\ee
As before, we will assume that this \emph{total} energy is somehow
available to \emph{each} of the two gases. The final ingredient is
the R-charge density $\jm=\jg$, which should be split in some way
between the two gases:
\be \label{jm}
\jm=\j+\bar{\j}~.
\ee
Given the form of (\ref{jg}), it is natural to propose that, at
equilibrium,
\be \label{jjbar}
\j={\pi^3\over 2\kappa^2}lr_0^4 e^{\alpha}, \qquad
\bar{\j}={\pi^3\over 2\kappa^2}lr_0^4 e^{-\alpha}.
\ee

Based on the AdS/CFT prediction (\ref{sgubser}) for the entropy of
a strongly-coupled charged SYM gas, the entropy of our microscopic
system is
\be \label{smcharged}
\sm=\frac{2^{5/4}3^{-3/4}\sqrt{\pi
N}e^{3/4}}{\sqrt{\sqrt{1+\chi}+\sqrt{\chi}}} +
\frac{2^{5/4}3^{-3/4}\sqrt{\pi
\bN}e^{3/4}}{\sqrt{\sqrt{1+\bar{\chi}}+\sqrt{\bar{\chi}}}}~,
\ee
where
\be \label{chichibar}
\chi={27\pi^2\j^4\over 8N^2e^3}~, \qquad
\bar{\chi}={27\pi^2\bar{\j}^4\over 8\bN^2e^3}~.
\ee
Using (\ref{nnbar}), (\ref{echarged}) and (\ref{jjbar}), one finds
that the ratios (\ref{chichibar}) simplify to
\be
\chi={l^4\over 4 r_0^4}=\bar{\chi}~,
\ee
and the two terms in (\ref{smcharged}) then combine into
\be \label{smchargedsimp}
\sm=\frac{2\pi^4\kappa^{-2}r_0^6\cosh\alpha}
{\sqrt{\sqrt{r_0^4+l^4/4}+l^2/2}}~.
\ee
Amazingly, multiplying the numerator and denominator by
$\sqrt{\sqrt{r_0^4+l^4/4}-l^2/2}$, we obtain perfect agreement
with the supergravity expression (\ref{sg})~! And, perhaps even
more remarkably, the assumed values (\ref{nnbar}) and
(\ref{jjbar}) can be verified to be precisely the ones that
maximize the entropy (\ref{smcharged}) for a fixed charge
(\ref{qm}) and mass (\ref{mmcharged}).

\section{Conclusions} \label{conclsec}

In this paper we have shown that the D3-\aD\ model of
\cite{beto1}, originally proposed as a microscopic description of
the non-rotating black threebrane, can also successfully account
for the entropy of the rotating threebrane arbitrarily far from
extremality (including in particular the neutral case, which
corresponds to a Kerr black hole in seven spacetime dimensions).
We regard this as a highly non-trivial test of the model.

For maximal conceptual clarity, it is worth commenting here on the
role played by the AdS/CFT correspondence in our calculations. In
the microscopic model, the entropy arises from two
strongly-coupled $\cN=4$ SYM gases. AdS/CFT is invoked as a tool
to determine this entropy, and the result, after a maximization
procedure, is then compared to the supergravity entropy. Since the
AdS/CFT prediction for the SYM entropy is \emph{deduced} from
supergravity, our calculation might at first glance appear to be
circular. The point to be emphasized, however, is that AdS/CFT
uses only the \emph{near-extremal} threebrane, whereas our
analysis extends arbitrarily far away from extremality. In
addition, the number of D3-branes in the model is not determined
uniquely by the charge of the system, but is instead fixed
thermodynamically. At the very least, then, the agreement found
here and in \cite{beto1} is a rather non-trivial property of the
supergravity formulae, implying a relation between the behavior of
the threebrane entropy in two completely different regimes.
Moreover, this property appears to have some degree of
universality, for it applies not only to the (rotating or not)
Type IIB threebrane, but also to the twobrane and fivebrane of
eleven-dimensional supergravity \cite{beto1}, and perhaps to other
systems as well \cite{hms,antientropy}. Even from this
perspective, then, the agreement would call for an explanation.

A particularly significant aspect of our results relates to what
in \cite{beto1} was regarded as a numerical discrepancy. As
explained in Section \ref{neutralsubsec} of the present paper, in
\cite{beto1} it was found that the entropy $\sm$ of the \DD\ model
was in perfect agreement with the functional form of the
supergravity entropy $\sg$; but, under the physically self-evident
condition that the total energy $e$ available to the two SYM gases
be \emph{split} between them, it was a factor of $2^{3/4}$ too
small, $\sg=2^{3/4}\sm$. Since $\sm\propto e^{3/4}$, one way to
phrase this disagreement was that the supergravity entropy behaved
as if each gas somehow had access to the \emph{total} energy
$e$.\footnote{The numerical discrepancies found in \cite{beto1}
for the twobrane and fivebrane cases were also of the form
$\sg=2^{p/(p+1)}\sm$, with $p$ the dimension of the brane, and
thus lent themselves to exactly the same interpretation.} In the
rotating case analyzed in this paper, the entropy of the model
depends on the energy $e$ of the gases not just through a simple
factor of $e^{3/4}$, but through the rather complicated function
seen in (\ref{smcharged}). It is quite remarkable then that,  to
reproduce the supergravity entropy, one again requires nothing
more and nothing less than the  assumption that \emph{each} gas
carries the \emph{total} energy $e$. Needless to say, the most
important pending task in connection with the brane-antibrane
model analyzed in \cite{beto1} and the present paper is to find a
physically reasonable explanation for this rather bizarre result.

As noted already in \cite{beto1}, it would also be interesting to
test the model by computing quantities other than the entropy.
Beyond that, there remain of course the ever-present questions
about the precise way in which the geometric and causal properties
of the black brane are encoded in the field-theoretic description.

\newpage
\noindent {\textbf{Note Added:}} After this paper was submitted to
the arXiv preprint server, we learned that overlapping results
have been independently obtained in \cite{ps}.

\section*{Acknowledgements}

This work was partially supported by Mexico's National Council of
Science and Technology (CONACyT), under grants CONACyT-U-40745-F and CONACyT-NSF E-120-0837.
The work of H.H. was supported by CONACyT Scholarship number 113375.


\begin{thebibliography}{99}

\bibitem{stringbh}
A. Strominger, C. Vafa, ``Microscopic origin of the
Bekenstein-Hawking entropy,'' Phys.\ Lett.\ B {\bf 379} (1996) 99,
[arXiv:hep-th/9601029]; \\
C.~G.~Callan and J.~M.~Maldacena, ``D-brane Approach to Black Hole
Quantum Mechanics,'' Nucl.\ Phys.\ B {\bf 472} (1996) 591,
[arXiv:hep-th/9602043]; \\
G.~T.~Horowitz and A.~Strominger, ``Counting States of
Near-Extremal Black Holes,'' Phys.\ Rev.\ Lett.\  {\bf 77} (1996)
2368, [arXiv:hep-th/9602051].

\bibitem{loopbh}
A.~Ashtekar, J.~Baez, A.~Corichi and K.~Krasnov, ``Quantum
geometry and black hole entropy,'' Phys.\ Rev.\ Lett.\  {\bf 80},
904 (1998) [arXiv:gr-qc/9710007]; \\
A.~Ashtekar, J.~C.~Baez and K.~Krasnov, ``Quantum geometry of
isolated horizons and black hole entropy,'' Adv.\ Theor.\ Math.\
Phys.\  {\bf 4}, 1 (2000) [arXiv:gr-qc/0005126];\\
A.~Ashtekar and A.~Corichi, ``Non-minimal couplings, quantum
geometry and black hole entropy,'' Class.\ Quant.\ Grav.\  {\bf
20}, 4473 (2003) [arXiv:gr-qc/0305082].

\bibitem{stringrev}
S.~R.~Das and S.~D.~Mathur, ``The Quantum Physics Of Black Holes:
Results From String Theory,'' Ann.\ Rev.\ Nucl.\ Part.\ Sci.\
{\bf 50} (2000) 153 [arXiv:gr-qc/0105063].

\bibitem{looprev}
A.~Ashtekar, ``Classical and Quantum Physics Of Isolated Horizons:
A Brief Overview,'' Lect.\ Notes Phys.\  {\bf 541}, 50 (2000).
[arXiv:gr-qc/9910101].

\bibitem{stringbhcorr}
M.~J.~Bowick, L.~Smolin and L.~C.~Wijewardhana, ``Does String
Theory Solve The Puzzles Of Black Hole Evaporation?,''
Gen.\ Rel.\ Grav.\  {\bf 19} (1987) 113; \\
L.~Susskind, ``Some speculations about black hole entropy in
string theory,''
[arXiv:hep-th/9309145]; \\
E.~Halyo, A.~Rajaraman and L.~Susskind, ``Braneless black holes,''
Phys.\ Lett.\ B {\bf 392} (1997) 319,
[arXiv:hep-th/9605112]; \\
E.~Halyo, B.~Kol, A.~Rajaraman and L.~Susskind, ``Counting
Schwarzschild and charged black holes,'' Phys.\ Lett.\ B {\bf 401}
(1997) 15,
[arXiv:hep-th/9609075]; \\
G.~T.~Horowitz and J.~Polchinski, ``Self gravitating fundamental
strings,'' Phys.\ Rev.\ D {\bf 57} (1998) 2557,
[arXiv:hep-th/9707170]; \\
R.~R.~Khuri, ``Entropy and string / black hole correspondence,''
Nucl.\ Phys.\ B {\bf 588} (2000) 253 [arXiv:hep-th/0006063].
T.~Damour, ``Quantum strings and black holes,''
arXiv:gr-qc/0104080;\\
E.~Halyo, ``Universal counting of black hole entropy by strings on
the stretched
horizon,'' JHEP {\bf 0112} (2001) 005 [arXiv:hep-th/0108167]; \\
A.~J.~M.~Medved, ``Asymptotically flat holography and strings on
the horizon,'' arXiv:hep-th/0202083.

\bibitem{hp}
G.~T.~Horowitz and J.~Polchinski, ``A correspondence principle for
black holes and strings,'' Phys.\ Rev.\ D {\bf 55} (1997) 6189,
[arXiv:hep-th/9612146].

\bibitem{bfss}
T.~Banks, W.~Fischler, S.~H.~Shenker and L.~Susskind, ``M theory
as a matrix model: A conjecture,'' Phys.\ Rev.\ D {\bf 55}, 5112
(1997), [arXiv:hep-th/9610043].

\bibitem{matrixbh}
T.~Banks, W.~Fischler, I.~R.~Klebanov and L.~Susskind,
``Schwarzschild black holes from matrix theory,'' Phys.\ Rev.\
Lett.\  {\bf 80} (1998) 226, [arXiv:hep-th/9709091];\\
I.~R.~Klebanov and L.~Susskind, ``Schwarzschild black holes in
various dimensions from matrix theory,'' Phys.\ Lett.\ B {\bf 416}
(1998) 62,
[arXiv:hep-th/9709108]; \\
G.~T.~Horowitz and E.~J.~Martinec, ``Comments on black holes in
matrix theory,'' Phys.\ Rev.\ D {\bf 57} (1998) 4935,
[arXiv:hep-th/9710217]; \\
E.~Halyo, ``Six dimensional Schwarzschild black holes in M(atrix)
theory,''
[arXiv:hep-th/9709225]; \\
M.~Li, ``Matrix Schwarzschild black holes in large N limit,'' JHEP
{\bf 9801} (1998) 009,
[arXiv:hep-th/9710226]; \\
T.~Banks, W.~Fischler, I.~R.~Klebanov and L.~Susskind,
``Schwarzschild black holes in matrix theory. II,'' JHEP {\bf
9801} (1998) 008,
[arXiv:hep-th/9711005]; \\
T.~Banks, W.~Fischler and I.~R.~Klebanov, ``Evaporation of
Schwarzschild black holes in matrix theory,'' Phys.\ Lett.\ B {\bf
423} (1998) 54,
[arXiv:hep-th/9712236]; \\
N.~Ohta and J.~G.~Zhou, ``Euclidean path integral, D0-branes and
Schwarzschild black holes in  matrix theory,''
Nucl.\ Phys.\ B {\bf 522} (1998) 125 [arXiv:hep-th/9801023]; \\
S.~Kalyana Rama, ``Matrix theory description of Schwarzschild
black holes in the regime $N >> S$,'' Phys.\ Rev.\ D {\bf 59}
(1999) 024006, [arXiv:hep-th/9806225].

\bibitem{boostbh}
S.~R.~Das, S.~D.~Mathur, S.~Kalyana Rama and P.~Ramadevi,
``Boosts, Schwarzschild black holes and absorption cross-sections
in M  theory,'' Nucl.\ Phys.\ B {\bf 527} (1998) 187,
[arXiv:hep-th/9711003];\\
F.~Englert and E.~Rabinovici, ``Statistical entropy of
Schwarzschild black holes,'' Phys.\ Lett.\ B {\bf 426} (1998) 269,
[arXiv:hep-th/9801048]; \\
R.~Argurio, F.~Englert and L.~Houart, ``Statistical entropy of the
four dimensional Schwarzschild black hole,'' Phys.\ Lett.\ B {\bf
426} (1998) 275, [arXiv:hep-th/9801053].

\bibitem{seiberg}
N.~Seiberg, ``Why is the matrix model correct?,'' Phys.\ Rev.\
Lett.\  {\bf 79} (1997) 3577 [arXiv:hep-th/9710009].

\bibitem{rindep}
A.~G\"uijosa, ``Is physics in the infinite momentum frame
independent of the compactification radius?,'' Nucl.\ Phys.\ B
{\bf 533} (1998) 406 [arXiv:hep-th/9804034].

\bibitem{malda}
J.~Maldacena, ``The large $N$ limit of superconformal field
theories and supergravity,'' Adv.\ Theor.\ Math.\ Phys.\  {\bf 2},
231 (1998) [Int.\ J.\ Theor.\ Phys.\  {\bf 38}, 1113 (1998)],
[arXiv:hep-th/9711200].

\bibitem{imsy}
N.~Itzhaki, J.~M.~Maldacena, J.~Sonnenschein and S.~Yankielowicz,
``Supergravity and the large N limit of theories with sixteen
supercharges,'' Phys.\ Rev.\ D {\bf 58} (1998) 046004
[arXiv:hep-th/9802042].

\bibitem{kll}
D.~Kabat, G.~Lifschytz and D.~A.~Lowe,
 ``Black hole thermodynamics from calculations in strongly coupled gauge
theory,'' Phys.\ Rev.\ Lett.\  {\bf 86} (2001) 1426 [Int.\ J.\
Mod.\ Phys.\
A {\bf 16} (2001) 856] [arXiv:hep-th/0007051];\\
D.~Kabat, G.~Lifschytz and D.~A.~Lowe, ``Black hole entropy from
non-perturbative gauge theory,'' Phys.\ Rev.\ D {\bf 64} (2001)
124015 [arXiv:hep-th/0105171];\\
N.~Iizuka, D.~Kabat, G.~Lifschytz and D.~A.~Lowe, ``Probing black
holes in non-perturbative gauge theory,'' Phys.\ Rev.\ D {\bf 65}
(2002) 024012 [arXiv:hep-th/0108006].

\bibitem{ikll}
N.~Iizuka, D.~Kabat, G.~Lifschytz and D.~A.~Lowe, ``Quasiparticle
picture of black holes and the entropy-area relation,'' Phys.\
Rev.\ D {\bf 67} (2003) 124001 [arXiv:hep-th/0212246];\\
N.~Iizuka, D.~Kabat, G.~Lifschytz and D.~A.~Lowe, ``Stretched
horizons, quasiparticles and quasinormal modes,'' Phys.\ Rev.\ D
{\bf 68} (2003) 084021 [arXiv:hep-th/0306209].

\bibitem{halyo}
E.~Halyo, ``Gravitational entropy and string bits on the stretched
horizon,'' arXiv:hep-th/0308166.

\bibitem{beto1}
U. Danielsson, A. G\"uijosa, M. Kruczenski, ``Brane-Antibrane
systems at finite temperature and the
 entropy of black holes,''
 JHEP {\bf 0109} (2001) 011, [arXiv:hep-th/0106201].

\bibitem{beto2}
U.~H.~Danielsson, A.~G\"uijosa and M.~Kruczenski, ``Black brane
entropy from brane-antibrane systems,'' Rev.\ Mex.\ F\'{\i}s.\
{\bf 49S2} (2003) 61 [arXiv:gr-qc/0204010].

\bibitem{hms}
G.~T.~Horowitz, J.~M.~Maldacena and A.~Strominger, ``Nonextremal
Black Hole Microstates and U-duality,'' Phys.\ Lett.\ B {\bf 383}
(1996) 151, [arXiv:hep-th/hep-th/9603109].

\bibitem{antientropy}
M.~Cvetic and D.~Youm, ``Entropy of Non-Extreme Charged Rotating
Black Holes in String Theory,'' Phys.\ Rev.\ D {\bf 54} (1996)
2612, [arXiv:hep-th/9603147];\\
G.~T.~Horowitz, D.~A.~Lowe and J.~M.~Maldacena, ``Statistical
Entropy of Nonextremal Four-Dimensional Black Holes and
U-Duality,'' Phys.\ Rev.\ Lett.\  {\bf 77} (1996) 430,
[arXiv:hep-th/9603195]; \\
R.~Kallosh and A.~Rajaraman, ``Brane-anti-brane Democracy,''
Phys.\ Rev.\ D {\bf 54} (1996) 6381,
[arXiv:hep-th/9604193]; \\
J.~Zhou, H.~J.~M\"uller-Kirsten, J.~Q.~Liang and F.~Zimmerschied,
``M-branes, anti-M-branes and non-extremal black holes,'' Nucl.\
Phys.\ B {\bf 487} (1997) 155
[arXiv:hep-th/9611146]; \\
M.~S.~Costa and M.~J.~Perry, ``Landau degeneracy and black hole
entropy,'' Nucl.\ Phys.\ B {\bf 520} (1998) 205,
[arXiv:hep-th/9712026].

\bibitem{ddbar}
A.~Sen, ``Stable non-BPS bound states of BPS D-branes,'' JHEP {\bf
9808} (1998) 010,
[arXiv:hep-th/9805019]; \\
``Tachyon condensation on the brane antibrane system,'' JHEP {\bf
9808}, 012 (1998),
[arXiv: hep-th/9805170]; \\
``Descent relations among bosonic D-branes,'' Int.\ J.\ Mod.\
Phys.\ A {\bf 14}, 4061 (1999),
[arXiv:hep-th/9902105]; \\
``Universality of the tachyon potential,'' JHEP {\bf 9912}, 027
(1999), [arXiv:hep-th/9911116].

\bibitem{krause}
A.~Krause, ``Schwarzschild black holes from brane-antibrane
pairs,'' arXiv:hep-th/0204206.

\bibitem{tempddbar}
W.~H.~Huang, ``Boundary string field theory approach to high
temperature tachyon
potential,'' arXiv:hep-th/0106002; \\
W.~H.~Huang, ``Brane-antibrane interaction under tachyon
condensation,'' Phys.\ Lett.\ B {\bf 561} (2003) 153
[arXiv:hep-th/0211127];\\
K.~Hotta, ``Brane-antibrane systems at finite temperature and
phase transition near the Hagedorn temperature,'' JHEP {\bf 0212}
(2002) 072
[arXiv:hep-th/0212063];\\
K.~Hotta, ``Finite temperature systems of brane-antibrane on a
torus,'' JHEP {\bf 0309} (2003) 002 [arXiv:hep-th/0303236].


\bibitem{gubser1}
S.~S.~Gubser, ``Thermodynamics of spinning D3-branes,'' Nucl.\
Phys.\ B {\bf 551} (1999) 667 [arXiv:hep-th/9810225].

\bibitem{klt}
P.~Kraus, F.~Larsen and S.~P.~Trivedi, ``The Coulomb branch of
gauge theory from rotating branes,'' JHEP {\bf 9903} (1999) 003
[arXiv:hep-th/9811120].

\bibitem{gkp}
S.~S.~Gubser, I.~R.~Klebanov and A.~W.~Peet, ``Entropy and
Temperature of Black 3-Branes,'' Phys.\ Rev.\ D {\bf 54} (1996)
3915, [arXiv:hep-th/9602135].

\bibitem{threequarters}
S.~S.~Gubser, I.~R.~Klebanov and A.~M.~Polyakov, ``Gauge theory
correlators from non-critical string theory,'' Phys.\ Lett.\ B
{\bf 428} (1998) 105,
[arXiv:hep-th/9802109]; \\
S.~S.~Gubser, I.~R.~Klebanov and A.~A.~Tseytlin, ``Coupling
constant dependence in the thermodynamics of $\cN = 4$
supersymmetric Yang-Mills theory,'' Nucl.\ Phys.\ B {\bf 534}
(1998) 202, [arXiv:hep-th/9805156].

\bibitem{ps}
A.~W.~Peet and O.~Saremi, to appear.

\end{thebibliography}
\end{document}